\documentclass[journal,draftcls,onecolumn,12pt,twoside]{IEEEtranTCOM}
\usepackage{amsopn, amsmath, fancybox, epsfig, amssymb, color, bm, dsfont , latexsym, graphicx,
accents, subfigure, algorithm, algpseudocode, multirow, caption, pifont, wasysym, stmaryrd}
\newcommand{\LRT}[2]{\overset{#1}{\underset{#2}{\gtrless}}}
\begin{document}

\title{Dynamic Weighted Bit-Flipping Decoding Algorithms for LDPC Codes }
\author{Tofar~C.-Y.~Chang,~\IEEEmembership{Student Member,~IEEE}~and~
        Yu~T.~Su,~\IEEEmembership{Senior Member,~IEEE}\\emails: \{tofar.cm96g, ytsu\}@nctu.edu.tw.%
        }
\markboth{IEEE Transactions on Communications}{Submitted paper}
\maketitle
\begin{abstract}
Bit-flipping (BF) decoding of low-density parity-check codes is of low complexity but gives inferior performance in general. To improve performance and provide new BF decoder options for complexity-performance tradeoffs, we propose new designs for the flipping function (FF), the flipped bit selection (FBS) rule and the checksum weight updating schedule. The new FF adjusts the checksum weights in every iteration while our FBS rules take more information into account. These two modifications represent efforts to track more closely the evolutions of both check and variable nodes' reliabilities. Two selective update schedules are proposed to offer more performance and complexity tradeoffs.

The combinations of the new FBS rule and known FFs result in new BF decoders with improved performance and a modest complexity increase. On the other hand, combining the new FF and FBS rule gives a new decoder with performance comparable to that of the normalized min-sum algorithm while if we use a much simpler FBS rule instead, the decoder suffers little performance loss with reduced complexity. We also present a simple decision-theoretical argument to justify the new checksum weight formula and a time-expanded factor graph model to explain the proposed selective weight-updating schedules.
\end{abstract}


\section{Introduction}
Low-density parity-check (LDPC) codes have been shown to asymptotically give near-capacity
performance when the sum-product algorithm (SPA) is used for decoding \cite{MacKayLDPC2}.
Gallager proposed two alternatives that use only hard-decision bits \cite{GallLDPC}.
These so-called bit-flipping (BF) algorithms flip one or a group of bits based on the
values of the flipping functions (FFs) computed in each iteration. The FF associated
with a variable node (VN) is a \emph{reliability metric} of the corresponding bit decision and
depends on the binary-valued checksums of the VN's connected check nodes (CNs). Although
BF algorithms are much simpler than the SPA, their performance is far from optimal. To reduce
the performance gap, many variants of Gallager's BF algorithms have been proposed. Most
of them tried to improve the VN's reliability metric (the FF) and/or the method of
selecting the flipped bits, achieving different degrees of bit error rate (BER) and
convergence rate performance enhancements at the cost of higher complexity.

The class of weighted bit-flipping (WBF) algorithms \cite{RyanLin}-\cite{RRWBF} assign proper
weights to the binary checksums. Each weight can be regarded as a reliability metric on the
corresponding checksum and is a function of the associated soft received channel values.
Another approach called gradient descent bit-flipping (GDBF) algorithm was proposed by
Wadayama {\it et al.} \cite{GDBF}. Instead of using a weighted checksum based FF, the GDBF
algorithm derives its FF by computing the gradient of a nonlinear objective function which
is equivalent to the log-likelihood function of the bit decisions with checksum constraints.
It was shown that the GDBF algorithm outperforms most known WBF algorithms when the VN
degree is small. Sundararajan {\it et al.} \cite{NGDBF} modified this FF by introducing a
weighting on syndrome and a zero-mean Gaussian perturbation term. The resulting noisy GDBF
(NGDBF) algorithm improves the performance of the GDBF algorithm which is further enhanced
by adding a re-decoding process \cite{RNGDBF}.

For the WBF algorithms, the weights are decided by the soft received channel values and
remain unchanged throughout the decoding process. Since the weights reflect the decoder's
belief on the checksums which, in turn, depend on those of the associated VNs' FF and bit
decisions, the associated checksum weights should be updated accordingly.
In \cite{RBF}, a reliability-based schedule is used in the initial decoding iteration to
forward only reliable VN and CN's messages. Nguyen and Vasi\'{c} \cite{TBBF} employed
an extra bit to adaptively represent the reliabilities of VN and CN messages and developed
a class of two-bit BF algorithms along with algorithm(s)-selection procedures. In this paper,
we present dynamic weighted BF (DWBF) algorithms that assign dynamic checksum weights which
are updated according to a nonlinear function of the associated VNs' FF values.
As we shall show, the nonlinear function has the effects of quarantining unreliable checksums
(which is similar to the method used in the first decoding iteration of \cite{RBF}) while
dynamically adjusts the more reliable checksums' weights. A simple decision theoretical
interpretation is given to explain the effect of the nonlinear action and justify the
threshold selection. We also suggest two selective weight-updating schedules which offer
additional performance-complexity trade-offs. A time-expanded factor graph model is used to
illustrate the weight-updating schedules.

The single-bit BF algorithms flip only the least reliable bit thus result in slow convergence
rates. For this reason, many a multiple-flipped-bit selection rule was suggested \cite{GDBF}--\cite{RNGDBF}.
By simultaneously flipping the selected bits, a BF decoder can offer rapid convergence but,
sometimes, at the expense of performance loss. A bit selection rule may consist of simple
threshold comparisons or include a number of steps involving different metrics. It is usually
designed assuming a specific FF is used and may not be suitable when a different FF or metric
is involved. Moreover, the FF value may not provide sufficient information for making a tentative
bit decision, we propose a new flipped bit selection (FBS) rule that takes into account both
the FF value and other information from related CNs.

FF, checksum weight computing, weight-updating schedules, and FBS rule are major constituent
parts of a DWBF decoder. Our proposals on these parts offer a variety of new design 
options and tradeoffs. The efficiencies of {\it using the proposed schemes jointly or separately with 
existing designs} are evaluated by examining the corresponding numerical error rate and convergence 
behaviors. We show that our single-bit DWBF algorithm provides significant performance improvement 
over the existing single-bit GDBF and WBF algorithms. Our FBS rule works very well with different
FFs and outperforms known FBS rules. Moreover, the selective weight-updating schedules suffer
little performance degradation while offer significant complexity reduction when the CN and
VN degrees are small.

Note that since the checksum weights are crucial parameters of a DWBF decoder's FF and their updates 
depend in turn on the FF values computed at the previous iteration, we henceforth mean both the FF and
the associated weight computing formula whenever FF is mentioned.

The rest of this paper is organized as follows. In Section \ref{section:Preliminaries}, we define
the basic system parameters and give a brief overview of various BF decoding algorithms, their FFs
and FBS rules. In Section \ref{section:ChkR}, we introduce a new FF and its checksum weight-updating
formula. A simple decision theoretical justification is given. We consider
single-bit BF decoders and present two weight-updating schedules as well as their graphic models
in Section \ref{section:single-DWBF}. The performance of the our single-bit DWBF algorithm and
some known single-bit BF algorithms are compared in the same section. We develop a new multi-bit
FBS rule and present the error rate and convergence behaviors of various multi-bit BF decoder
structures based on the new FBS rule in Section \ref{section:multi-DWBF}. These decoders'
complexities are analyzed in details for evaluating various performance and complexity tradeoffs.
Finally, conclusion remarks are drawn in Section \ref{section:conclusion}.

\section{Background and Related Works}\label{section:Preliminaries}
\subsection{Notations and the Basic Algorithm}\label{subsection:Basics}
We denote by $(N,K)(d_v,d_c)$ a regular binary LDPC code $\mathcal{C}$
with VN degree $d_v$ and CN degree $d_c$, i.e.,
$\mathcal{C}$ is the null space of an $M\times N$ parity check matrix $\bm{H}=[H_{mn}]$
which has $d_v$ 1's in each column and $d_c$ 1's in each row.
Let $\bm{u}$ be a codeword of $\mathcal{C}$
and assume that the BPSK modulation is used so that
a codeword $\bm{u}= (u_0,u_1,\cdots,u_{N-1})$, $u_i\in\{0,1\}$, is mapped into
a bipolar sequence $\bm{x}=(x_0, x_1,\cdots,x_{N-1})=(1-2u_0,
1-2u_1,\cdots,1-2u_{N-1})$ for transmission. The equivalent baseband transmission
channel is a binary-input Gaussian-output channel characterized by additive
zero-mean white Gaussian noise with two-sided power spectral density of $N_0/2$ W/Hz.
Let $\bm{y}=(y_0,y_1,\cdots, y_{N-1})$ be the sequence of soft channel values obtained at
the receiver's coherent matched filter output. The sequence $\bm{z}=(z_0,z_1,\cdots,z_{N-1})$,
where $z_i\in\{0,1\}$, is obtained by taking hard-decision on each
components of $\bm{y}$. Let $\hat{\bm{u}}=(\hat{u}_0,\hat{u}_1,\cdots, \hat{u}_{N-1})$
be the tentative decoded binary sequence at the end of a BF decoding iteration.
We compute the syndrome (checksum) vector $\bm{s}=(s_0,s_1,\cdots,s_{M-1})$ by
$\bm{s}=\hat{\bm{u}}\cdot\bm{H}^{T}$(mod 2). We further denote the $n$th VN by $v_n$,
the set of indices of its connecting CNs by $\mathcal{M}(n)$,
and the set of indices of the VNs checked by
the $m$th CN $c_m$ by $\mathcal{N}$($m$). The indices of CNs in $\mathcal{M}(n)$
are determined by the nonzero elements of the $n$th column of $\bm{H}$ whereas those
in $\mathcal{N}$($m$) are by the $m$th row of $\bm{H}$.

A generic BF decoding algorithm can be described by {\bf Algorithm} \ref{alg:BF} below
which involves four important parameters: $l$, the iteration number; $l_{\text{max}}$,
the maximum iteration number; $E_n$, the FF; and $\mathcal{B}$, the index set of the
flipped bits, or the flipped bit (FB) set for short. This algorithm performs two basic
tasks:
1) computing $E_n$'s (\textbf{Step} 2) and
2) generating the FB set $\mathcal{B}$ (\textbf{Step} 3).
Most earlier works focused on improving either 1) or 2).
\begin{algorithm}
    \caption{Bit-Flipping Decoding Algorithms}\label{alg:BF}
    \textbf{Initialization} Set $l=0$ and $\hat{\bm{u}}=\bm{z}$.\\ 
    \textbf{Step} 1 $\forall ~m \in O_M\triangleq\{0,1,\ldots,M-1\}$, compute
        \begin{eqnarray}\label{eqn:syn}
            s_m=\sum_{n\in\mathcal{N}(m)}\hat{u}_n H_{mn} \text{(mod 2)}.
        \end{eqnarray}\par
    \hskip\algorithmicindent\hskip\algorithmicindent If $\bm{s}=\bm{0}$ or $l=l_{\text{max}}$, stop decoding and output $\hat{\bm{u}}$; otherwise, $l\leftarrow l+1$.\\
    \textbf{Step} 2 $\forall~n \in O_N\triangleq\{0,1,\ldots,N-1\}$, compute the FF $E_n$.\\
    \textbf{Step} 3 Use the FFs obtained in \textbf{Step} 2 to update the flipped bit set $\mathcal{B}$. \\
    \textbf{Step} 4 Flip $\hat{u}_{n}$ for all $n\in\mathcal{B}$ and go to \textbf{Step} 1.
\end{algorithm}
An FF, sometimes referred to as {\it cost function} or {\it inversion function} \cite{GDBF},
is used as a reliability metric on a VN's tentative decision. Given the FF values and the
FBS rule, we select a set of VNs and flip the corresponding
tentative decisions (bits). Choosing the most unreliable bits or the bits whose FF values
exceed a threshold are the two most popular rules. For the former rule, usually only one
bit is flipped if the soft-valued channel information is employed in the FF, resulting in
slow convergence. By contrast, the latter rule often gives faster convergence rate but
possibly at the cost of performance loss. We briefly review the known FFs and FBS rules
in the following paragraphs.

\subsection{Flipping Functions of BF Decoding Algorithms}\label{subsection:oldFFs}
Gallager proposed that a simple sum of binary checksums be used as the FF \cite{GallLDPC}
\begin{eqnarray}\label{eqn:FF-Gallager}
    E_n=-\sum_{m\in\mathcal{M}(n)}(1-2s_m).
\end{eqnarray}
(\ref{eqn:FF-Gallager}) implies that the FF value is inversely proportional to the bit
decision reliability as it is an increasing function of the number of nonzero checksums
(i.e., unsatisfied check nodes, UCNs).

By taking into account soft-valued channel information and assigning checksum weights,
later modifications of Gallager's FF can be described by the following general formula
\begin{eqnarray}\label{eqn:FF-general}
    E_n=-\alpha_1\cdot\phi(\hat{u}_n,y_n)-\sum_{m\in\mathcal{M}(n)}w_{mn}(1-2s_m),
\end{eqnarray}
where $\alpha_1> 0$, $\phi(\hat{u}_n,y_n)$ is a reliability metric involving channel
value and/or bit decision, and, to be consistent with (\ref{eqn:FF-Gallager}), $w_{mn}\geq 0$.

For the WBF algorithm \cite{WBF}, $\phi(\hat{u}_n,y_n)=0$ and $w_{mn}$ is
\begin{eqnarray}\label{eqn:W-WBF}
    w_{mn}=\min_{n^{'}\in\mathcal{N}(m)}|y_{n^{'}}|,
\end{eqnarray}
The modified WBF (MWBF) algorithm \cite{MWBF} has
$\phi(\hat{u}_n,y_n)=|y_n|$ while the improved MWBF (IMWBF) algorithm \cite{IMWBF}
uses the same $\phi(\hat{u}_n,y_n)$ but replaces the checksum weight by
\begin{eqnarray}\label{eqn:W-IMWBF}
    w_{mn}=\min_{n^{'}\in\mathcal{N}(m)\backslash n}|y_{n^{'}}|
\end{eqnarray}
for the message passed from $c_m$ to $v_n$ should exclude that originated from $v_n$. For
the reliability ratio based WBF (RRWBF) algorithm \cite{RRWBF}, $\phi(\hat{u}_n,y_n)=0$ and
\begin{eqnarray}\label{eqn:W-RRWBF}
    w_{mn}=1/w^{'}_{mn}=\left(\beta\frac{|y_n|}{\max_{n'\in\mathcal{N}(m)}|y_{n'}|}\right)^{-1},
\end{eqnarray}
where $\beta$ is the normalizing factor to ensure that $\sum_{n\in\mathcal{N}(m)} w^{'}_{mn}=1$.

The GDBF algorithm of Wadayama \emph{et al.} \cite{GDBF} applies
the gradient descent method to minimize
\begin{eqnarray}\label{eqn:F(x)-GDBF}
    E(\hat{\bm{u}})=-\sum_{n=0}^{N-1}y_n(1-2\hat{u}_n)-\sum_{m=0}^{M-1}(1-2s_m)
\end{eqnarray}
with respect to $(1-2\hat{u}_n)$ and obtains the FF
\begin{eqnarray}\label{eqn:FF-GDBF}
    E_n=-y_n(1-2\hat{u}_n)-\sum_{m\in\mathcal{M}(n)}(1-2s_m),
\end{eqnarray}
which is equivalent to assigning $\alpha_1=1$, $\phi(\hat{u}_n,y_n)=y_n(1-2\hat{u}_n)$,
and $w_{mn}=1$ in (\ref{eqn:FF-general}).
Recently, Sundararajan \emph{et al.}\cite{NGDBF} introduced the so-called noisy GDBF (NGDBF)
algorithm based on
\begin{eqnarray}\label{eqn:FF-NGDBF}
    E_n=-y_n(1-2\hat{u}_n)-w\sum_{m\in\mathcal{M}(n)}(1-2s_m)+q_n,
\end{eqnarray}
where $q_n$'s in (\ref{eqn:FF-NGDBF}) are i.i.d. zero-mean Gaussian random perturbation
with a signal-to-noise ratio (SNR) dependent variance and $w$ is a constant syndrome weight.

For the above two FFs, (\ref{eqn:FF-GDBF}) and (\ref{eqn:FF-NGDBF}), $\phi(\hat{u}_n,y_n)
=y_n(1-2\hat{u}_n)$ is equal to $|y_n|$ when the bit decision $\hat{u}_n$ is the same as
the $y_n$-based hard decision, $z_n$; otherwise, its value is the same as $-|y_n|$.
This is consistent with the intuition that $\hat{u}_n=z_n$ implies that $\hat{u}_n$ is
likely to be correct and since $-E_n$ is a VN reliability metric, a positive
$\phi(\hat{u}_n,y_n)$ helps increasing $-E_n$ and preventing the $\hat{u}_n$ from being
flipped. In contrast, $\phi(\hat{u}_n,y_n)$ is always positive ($|y_n|$) in MWBF algorithms,
which means that the MWBF algorithms tend to trust $\hat{u}_n$ in spite of other evidence.

\subsection{Flipped Bit Selection Rules}\label{subsection:FlippedBitSet}
For the algorithms mentioned in the Section \ref{subsection:oldFFs},
only the bit(s) related to the VN having the largest FF value $E_n$ is (are) flipped in
each iteration, i.e., the FB set is
\begin{equation}\label{FBS-Single}
    \mathcal{B}=\{n|n=\arg\max_i E_i\}.
\end{equation}
As mentioned before, $|\mathcal{B}|=1$ and the corresponding convergence is often
very slow if $E_n$ has a soft-valued information term.

Flipping several bits in each iteration simultaneously can improve the convergence speed.
The simplest FBS rule for multi-bit BF decoding uses the FB set
\begin{equation}\label{FBS-Multi}
    \mathcal{B}=\{n|E_n\geq\Delta\},
\end{equation}
where the threshold $\Delta$ can be a constant or be adaptive. The optimal adaptive threshold
was derived by Gallager \cite{GallLDPC}, assuming that no cycle appears in the code graph.
Since practical finite-length LDPC codes usually have cycles and the optimal thresholds can only be
found through time-consuming simulations, two ad-hoc methods which automatically adjust $\Delta$ were
suggested in \cite{ATBF} and \cite{AWBF}.
In the adaptive threshold BF (ATBF) algorithm \cite{ATBF}, the initial $\Delta$ is
found by simulation and subsequent thresholds are a monotonically non-increasing
function of the decoding iterations. The adaptive MWBF (AMWBF) algorithm \cite{AWBF}
adjusts the threshold by
\begin{eqnarray}\label{eqn:Thr-AWMBF}
    \Delta=E^*-|E^*|\left[1-\frac{w_\text{H}(\bm{s})}{M}\right],
\end{eqnarray}
where $E^*=\max_nE_n$ and $w_\text{H}(\bm{s})$ is the Hamming weight of the syndrome vector $\bm{s}$.

Sometimes, a tentative decoded vector $\hat{\bm{u}}$ may reappear several times during the decoding
process and form a decoding \emph{loop}. This may be caused, for example, by the event that an even number
of bits associated with a CN are flipped, leading to an unchanged checksum and then oscillating bit
decisions. To eliminate the occurrence of loops, the AMWBF algorithm includes the loop detection scheme
of \cite{AMBF} in its FBS rule so that if a loop is detected, the most reliable bit in $\mathcal{B}$
is removed. The parallel weighted BF (PWBF) algorithm \cite{PWBF} tries to reduce the loop occurrence
probability by having every UCN ($s_m=1$) send a constant \emph{flip signal} (FS) to
its least reliable linked VN (based on the FF of the IMWBF algorithm) and flips the bits in
\begin{eqnarray}\label{FBS-PWBF}
    \mathcal{B}=\{n|F_n\geq\Delta_{\text{FS}}\},
\end{eqnarray}
where $\Delta_{\text{FS}}$ is a constant optimized by simulations,
\begin{eqnarray}\label{eqn:FScnt}
    F_n=\sum_{m\in\mathcal{M}(n)}q_{mn}s_m,
\end{eqnarray}
and $q_{mn}$ is given by
\begin{eqnarray}\label{eqn:rp_mn}
	q_{mn}=\left\{
	\begin{array}{ll}
    	1,&\quad n=\arg\displaystyle\max_{n^{'}\in\mathcal{N}(m)}E_{n^{'}}\\
        0,&\quad \text{otherwise}\\
    \end{array}
    \right..
\end{eqnarray}
Since the above remedy can only eliminate loops with a certain probability, the improved
PWBF (IPWBF) algorithm employs the loop detection scheme of \cite{AMBF} and when a loop
is detected, it removes the bit(s) receiving the smallest $F_n$ from $\mathcal{B}$. This
algorithm also adds a bootstrapping step and a delay-handling procedure to further improve
the bit selection accuracy but achieves limited improvement for the codes with high column
degrees such as Euclidean geometry (EG) LDPC codes.

A hybrid GDBF (HGDBF) algorithm was proposed in \cite{GDBF}. In this algorithm, single-
and multi-bit BF decoding is performed alternatively and an \emph{escape process} is used
for preventing the decoding process from being trapped in local minima/loops. Two extensions
of the HGDBF algorithm were considered in \cite{MGDBF1} and \cite{MGDBF2} which require less
complexity at the expense of inferior performance. A multi-bit GDBF algorithm with a
probabilistic FBS rule was also suggested in \cite{PGDBF} for hard-decision decoding. With
the FF (\ref{eqn:FF-NGDBF}) and FBS rule of \cite{MGDBF2}, the multi-bit NGDBF (M-NGDBF)
algorithm \cite{NGDBF} achieves the same BER performance as that of the HGDBF decoder with
much less decoding iterations. However, \cite{RNGDBF} found that new trapping set conditions
may exist in the M-NGDBF decoder but can be eliminated by re-decoding with a different
perturbation sequence.


\section{Checksum Reliability and Dynamic Weights}\label{section:ChkR}
\subsection{BF Decoding and Checksum Weights}\label{subsection:ChkR}
In line with the belief propagation (BP) based SPA, $E_n$ is similar to the total
log-likelihood ratio (LLR) of $v_n$ and $-w_{mn}(1-2s_m)$ in (\ref{eqn:FF-general})
is analogous to the belief sent to $v_n$ by $c_m$. Unlike SPA, however,
for (\ref{eqn:W-WBF})--(\ref{eqn:W-RRWBF}), the latter remains unchanged unless an
$\hat{u}_n,~n \in ~\mathcal{N}(m)$ has been flipped, which leads to just a sign change
of the belief. The general FF format, (\ref{eqn:FF-general}), includes two major terms
that represent the decoder's confidence on a VN's tentative decision based respectively
on its channel value (or the correlation of the channel value and tentative decision)
and the reliabilities of the related checksums. Since the channel values remain fixed,
the checksums should be given adjusted weights at least in the later iterations when
the reliabilities of checksums change.

Although the flipping operation changes the reliability metric of $\hat{u}_n$ and the
related checksums, all the FFs used in known BF decoders use static $w_{mn}$ thereby
can neither reflect the dynamic of VNs' message passing nor offer self-adjustment
capability in accurately updating bit reliability information. We present dynamic
weight generation method in this section.


\subsection{Flipping Function and Decision Reliability}\label{subsection:newFF}
The review on BF algorithms in Section \ref{section:Preliminaries} indicates clearly that the FF value is a proper explicit
or implicit reliability metric of a VN's decision. As a checksum in turn is a function of the
associated VNs' decisions, the corresponding checksum weights should be updated according to the
current FF values. A reasonable candidate checksum weight is therefore given by
\begin{eqnarray}\label{eqn:dynWeight-pre}
    r_{mn}^{(l)}=\min_{n^{'}\in\mathcal{N}(m)\backslash n}-E_{n'}^{(l)},
\end{eqnarray}
where $E_n^{(l)}$ is the FF value of $v_n$ in the $l$th iteration.
However, (\ref{eqn:dynWeight-pre}) may result in negative weights which is inconsistent with
(\ref{eqn:FF-Gallager}) and (\ref{eqn:FF-general});
both are increasing functions of the nonzero checksum number.
To have proper positive checksum weights based on $E_n$, we consider the
likelihood ratio
\begin{eqnarray}\label{eqn:seperator}
    \Lambda(E_n)=\frac{f(E_n|H_0)}{f(E_n|H_1)},
\end{eqnarray}
where $H_0$ and $H_1$ denote the hypotheses that $\hat{u}_n=u_n$ and $\hat{u}_n\neq u_n$,
respectively. The conditional probability density function (pdf), $f(E_n|H_i)$, is the pdf of
those $E_n$'s associated with a correct or incorrect tentative bit decision at a given iteration.
It is to be interpreted as a conditional pdf averaged over all VNs. The basic decision theory
tells us that the optimal decision rule is given by
\begin{eqnarray}\label{eqn:Bayesian}
\Lambda(E_n) \LRT{H_0}{H_1} \frac{\pi_1(C_{01}-C_{11})}{\pi_0(C_{10}-C_{00})}
\end{eqnarray}
where $\pi_i=$ $P_r(H_i$ is true$)$ and $C_{ij}$ is the ``cost" for accepting $H_i$ while
$H_j$ is true. Unlike the conventional Bayesian minimizing error probability setting, both
the costs and the a priori probabilities are difficult to assess. For a BF decoder, a
tentative decision, except for the initial iteration, is determined by the previous decision
and the flipping decision. On the other hand, both conditional pdfs in (\ref{eqn:seperator})
depend on the definition of $E_n$ and the FBS rule. They vary from one iteration to another.
Although the dependence between $\hat{u}_n$ and $E_n$ is implicit as other parameters
are also intertwined and the conditional pdfs are thus difficult if not impossible to derive
analytically, they can be estimated numerically by simulations.
\begin{figure}
    \begin{center}
        \epsfxsize=3.7in
        \epsffile{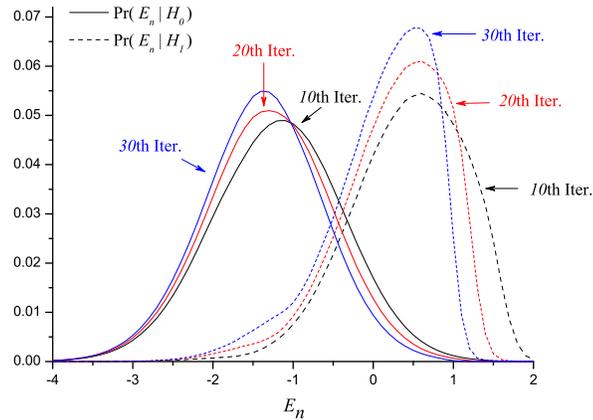}
        \caption{\label{fig:En-IMF}Conditional FF distributions for the IMWBF algorithm in decoding
        MacKay (816,272)(4,6) LDPC code (816.1A4.845 \cite{MacKayWeb}), SNR ($E_b/N_0$)= 4 dB.}
    \end{center}
\end{figure}
Assuming that the all-zero codeword is transmitted in every frame, i.e., $\hat{u}_n=0$ is
the correct decision, we depict in Fig. \ref{fig:En-IMF} the evolutions of both simulated
conditional pdfs with all related factors considered and averaged--for the IMWBF algorithm with
$E_n$ being the updated FF values at, say the $l$th iteration, after the tentative decision at
the $(l-1)$th iteration $\hat{u}_n$ was made. This figure indicates that $E_n$ does reflect the
correctness of the corresponding tentative decision in that a smaller (negative) FF value is
more likely to be associated with a correct decision. The probability of correct inference on
$\hat{u}_n$ based on $E_n$ depends on the separation (distance) between two pdfs. The figure,
however, demonstrates that the separation does not improve as the decoding iteration increases.
We propose a new FF in the next subsection and show in Section \ref{section:single-DWBF} that
this new FF is capable of overcoming the shortcoming of the IMWBF algorithm.

\subsection{A New Flipping Function}
Although $\pi_i$ and $C_{ij}$ are difficult to assess but the optimal Bayesian test, (\ref{eqn:Bayesian}),
can be simplified to a threshold test based on $E_n$. Given a suitable threshold $\eta$, Fig. \ref{fig:En-IMF}
indicates that if $-E_n > \eta$, the decoder is more likely to have made a correct bit decision $\hat{u}_n$
whose reliability is proportional to $-E_n$; otherwise, the decision is probably incorrect.

The foregoing discussion and the aim to have a weight-updating rule reflecting a more
accurate relation among bit decision, FF, and checksums, as elaborated in more details
below, suggest that we modify (\ref{eqn:dynWeight-pre}) as
\begin{eqnarray}\label{eqn:dynWeight}
    r_{mn}^{(l)}=\Omega\left(\min_{n^{'}\in\mathcal{N}(m)\backslash n}-E_{n'}^{(l)}\right)
    =\min_{n^{'}\in\mathcal{N}(m)\backslash n}\Omega(-E_{n'}^{(l)}),
\label{r_mn}
\end{eqnarray}

where
\begin{eqnarray}\label{eqn:clipper}
	\Omega(x)=\left\{
	\begin{array}{ll}
    	x-\eta,&\quad x \geq \eta\\
        0,&\quad x < \eta
    \end{array}
    \right..
\end{eqnarray}
The clipping operator, $\Omega(x)$, besides ensuring only positive weights are used,
can be interpreted as the decision for a CN to send no message to other linked VNs
when the associated FF values fail to exceed the threshold, which bears the flavor
of ``stop-and-go" algorithms that pass a CN-to-VN message only if it is deemed
reliable. Note that a checksum $s_m$ is determined by $d_c$ bit decisions, and if
$s_m=0$ and there is only one unreliable decision $\hat{u}_n$ (-$E_n<\eta$) among VNs
in $\mathcal{N}(m)$, the checksum is likely to be valid and the decision is in fact
correct. Hence $c_m$ should modify $E_n$ to increase $\Lambda(E_n)$ but not pass the
message $-r_{mn'}^{(l)}(1-2s_m)$ to other connected (reliable) VNs ($n'\in\mathcal{N}(m)
\setminus n$). In doing so, $E_n$ has a local (among $\mathcal{N}(m)$) maximum
FF decrease and the probability of reversing the bit decision is reduced. On the other
hand, if $s_m=1$, $\hat{u}_n$ is likely to be only local incorrect decision, the above
rule will result in a local maximum FF increase and thus a higher probability of
being flipped. When more than one $-E_n,~n \in \mathcal{N}(m)$ are clipped, no
message is sent from $c_m$ as the checksum itself is unreliable. The temporary
suspension of some message propagation induced by $\Omega(x)$ also has the desired
effect of containing the damage an incorrect message may have done and preventing the
decoder from being trapped in a local minimum. The above discourse confirms that
(\ref{eqn:clipper}) does fulfill the goal that the weight updating should have FFs,
checksums, and bit decisions join a cohesive effort in improving the performance of
a BF decoder.

The FF defined by (\ref{eqn:FF-general}) using the recursive weight-updating rule
(\ref{eqn:dynWeight}) tends to make the {\it check reliability} part of the FF,
$-\sum_{m\in\mathcal{M}(n)}r_{mn}(1-2s_m)$, starts to grow exponentially after
most of the correctable bits were flipped and the number of UCNs decreases to
just a few. It is conceivable that these reliability estimates should be given
different weight with more remote estimates having less weights. This can be done
by having the check reliabilities multiplied by a forgetting factor, $0<\alpha_2<1$,
as can be found in many recursive adaptive filters \cite{Haykin}. We are unable to
determine the optimal clipping threshold $\eta$ since a closed-form expressions for
$\text{Pr}(E_n|H_i)$ are practically unobtainable for reasons mentioned before.
Some simulation efforts, however, indicate that $\eta$ is close to 0 for several
BF decoders, independent of SNR and the iteration of interest. With the above ideas in
mind, we consider a new FF based on (\ref{eqn:dynWeight}) and (\ref{eqn:clipper})
using $\eta=0$:
\begin{eqnarray}\label{eqn:FF-DWBF}
  E_n^{(l)}=-y_n(1-2\hat{u}_n)-\alpha_2\sum_{m\in\mathcal{M}(n)}r_{mn}^{(l-1)}(1-2s_m),
\end{eqnarray}
where $0<\alpha_2<1$ is a positive damping (forgetting) constant to be optimized by
numerical experiments.

\section{New Single-Bit BF Decoding Algorithms}\label{section:single-DWBF}
We define a (checksum) weight-updating schedule as a rule that selects a subset $\mathcal{G}$
of $O_M$, which represents the set of CN indices, and updates only those $r_{mn}^{(l)}, m \in \mathcal{G}$.
Such a rule determines the message-passing paths in the decoding process
(see Fig. \ref{fig:check_update} below) hence is called a schedule.
When the updated CN index set $\mathcal{G}=O_M$, we call the schedule as the \emph{full weight-updating
schedule} (FWUS). Alternate schedules with $\mathcal{G}\neq O_M$ provide trade-offs between
computational complexity and error-rate performance. In this section, we introduce a class of
{\it single-bit dynamic weighted BF} decoding methods based on (\ref{eqn:dynWeight}), (\ref{eqn:FF-DWBF}),
and different weight-updating schedules.

\subsection{Single-Bit DWBF Decoding and Weight-Updating Schedules}\label{subsection:sDWBF_CNupdate}

Combining (\ref{eqn:dynWeight}) and (\ref{eqn:FF-DWBF}) with the consideration of
the choice of a weight-updating schedule,
we obtain the class of DWBF algorithms or, for simplicity, \textbf{Algorithm} \ref{alg:DWBF}.
\begin{algorithm}[H]
    \caption{DWBF Decoding Algorithm}\label{alg:DWBF}
    \textbf{Initialization} Set $l=0$, $\hat{\bm{u}}=\bm{z}$, and $\mathcal{G}=\mathcal{B}=\emptyset$.
    Initialize $r_{mn}^{(l)}$ by (\ref{eqn:W-IMWBF})
    for all $n\in\mathcal{N}(m),~m \in O_M$. Set $E_n^{(l)}=-y_n(1-2\hat{u}_n)$ for all $n\in O_N$ .\\
    \textbf{Step} 1 Compute $s_m$ for all $m \in O_M$. If $\bm{s}=\bm{0}$ or $l=l_{\text{max}}$, stop decoding and output $\hat{\bm{u}}$;
    otherwise, \par
    \hskip\algorithmicindent\hskip\algorithmicindent $l\leftarrow l+1$.\\
    \textbf{Step} 2 $\forall~ n \in O_N$, compute $E_n^{(l)}$ by $(\ref{eqn:FF-DWBF})$.\\
    \textbf{Step} 3 Update $\mathcal{B}$ and $\forall n\in\mathcal{B}$, flip $\hat{u}_{n}$ and
    $E_n^{(l)}\leftarrow -E_n^{(l)}$.\\
    \textbf{Step} 4 Update $\mathcal{G}$. Then, update $r_{mn}^{(l)}$ by (\ref{eqn:dynWeight})
    $\forall~ n\in\mathcal{N}(m),m\in\mathcal{G}$ and set $r_{mn}^{(l)}\leftarrow r_{mn}^{(l-1)}$ \par 
    \hskip\algorithmicindent\hskip\algorithmicindent $\forall~ n\in\mathcal{N}(m),~m\in O_M\backslash\mathcal{G}$ and
    go to \textbf{Step} 1.
\end{algorithm}

Algorithm 2 describes a general class of DWBF algorithms. For hard-decision decoding, $-y_n(1-2\hat{u}_n)$
in (\ref{eqn:FF-DWBF}) is replaced by $-(1-2z_n)(1-2\hat{u}_n)$ and $r_{mn}^{(l)}$ is initialized as $1$.
The FB set $\mathcal{B}$ in {\bf Step} 3 is determined by the FBS rule used, it can be (\ref{FBS-Single}),
{\bf Algorithm} 3 or 4 presented in Section \ref{section:multi-DWBF}. When (\ref{FBS-Single}) is used,
{\bf Algorithm} 2 is a single-bit DWBF algorithm and becomes a multi-bit DWBF algorithm if {\bf Algorithm}
3 or 4 is used as the FBS rule. In this section, we consider only the FBS rule (\ref{FBS-Single}), which
implies that only one bit is flipped (i.e., $|\mathcal{B}|=1$) at each iteration unless it is used for
hard-decision decoding. Hence the resulting decoder is referred to as the single-bit dynamic weighted BF
(S-DWBF) decoder.

As most FF values will change because of the recursive nature of (\ref{eqn:dynWeight}) and (\ref{eqn:FF-DWBF})
we may need to perform the FWUS in \textbf{Step} 4. This is one of the prices we have to pay when the dynamic
weights instead of the conventional constant weights are assigned to the checksums. We call the FWUS-based
single-bit decoding algorithm as the S-DWBF-F algorithm for simplicity. To lessen the computing load of this
algorithm, we reduce the size of $\mathcal{G}$ by prioritizing the CNs and update only those with a higher
priority.

We first notice that, to ensure that the newest updated information be broadcasted,
the weights of the flipped bits' linked checksums should have the highest updating priority.
Furthermore, for those VNs whose FF values change from one side of the clipping threshold $\eta$
of (\ref{eqn:clipper}) to the other side and undergo a \emph{reliability inversion}, their related
checksum weights should be renewed as well. With these considerations,
the \emph{selective weight-updating schedule A} (SWUS-A) updates only those checksums (CNs) whose
indices lie in
\begin{eqnarray} \nonumber
\mathcal{G}_{\text{A}}^{(l)}&\triangleq&\{m|m\in\mathcal{M}(n), n\in\mathcal{B}\}\\ \label{eqn:SWUS-A}
    &&\cup\{m|m\in\mathcal{M}(n), (-E_n^{(l)}-\eta)(-E_n^{(l-1)}-\eta)<0, n=0,1,\ldots,N-1\}.
\end{eqnarray}

\begin{figure}
    \begin{center}
        \subfigure[Selective weight-updating schedule A. \label{fig:update_A}]{
            \epsfxsize=3.4in
            \epsffile{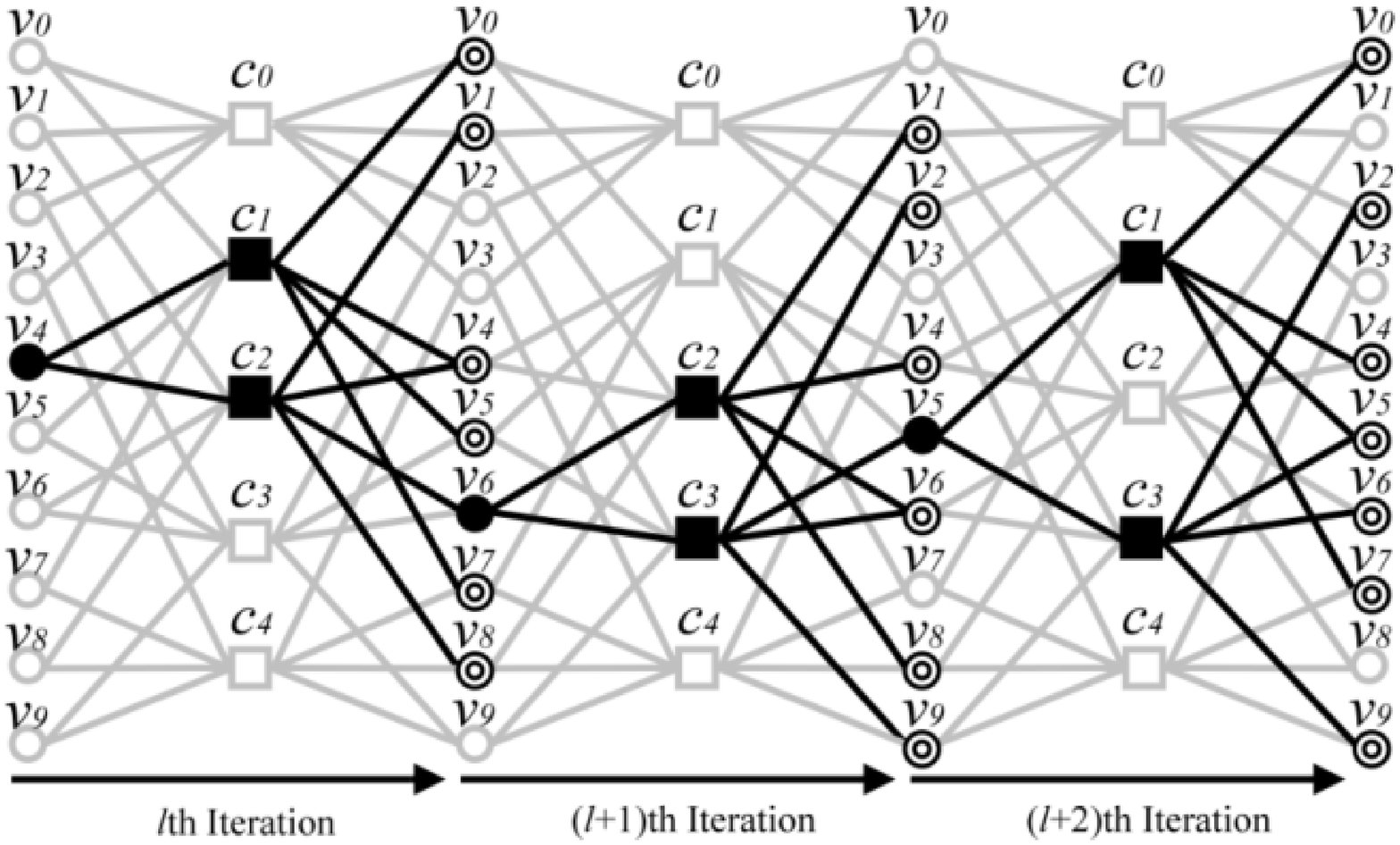}}
        \subfigure[Selective weight-updating schedule B. \label{fig:update_B}]{
            \epsfxsize=3.4in
            \epsffile{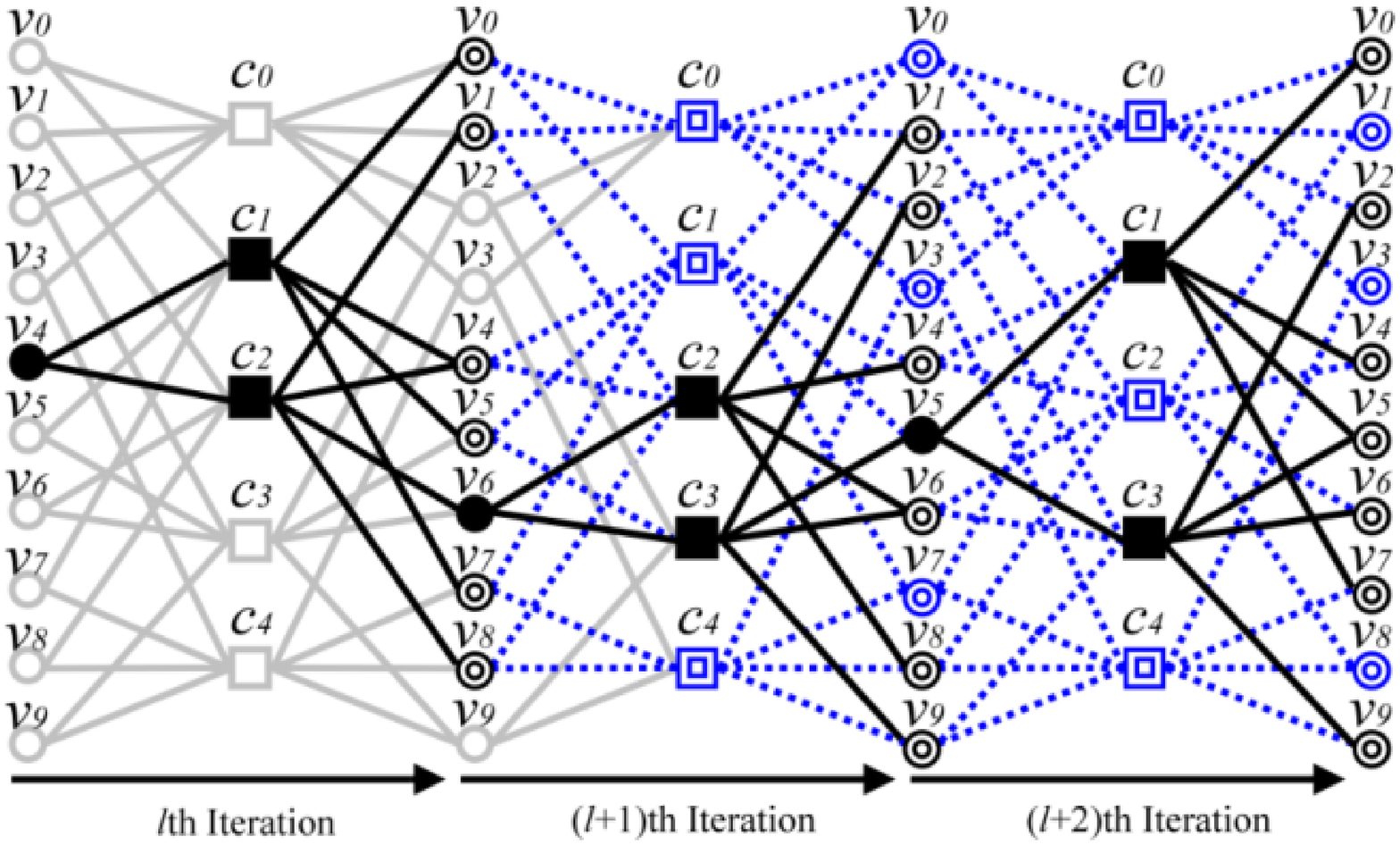}}
        \caption{\label{fig:check_update}The TEFGs for
        two different selective weight-updating schedules.}
    \end{center}
\end{figure}
The time-expanded factor graph (TEFG) shown in Fig. \ref{fig:update_A} is a simple example
illustrating how SWUS-A behaves, assuming that the only VN which generates
$\mathcal{G}_{\text{A}}^{(l)}=\{1,2\}$ is $v_4$. We denote this VN by \ding{108}, the CNs
\emph{visited} (selected) by the schedule by \ding{110}, and the VNs which receive new CN
messages by $\varocircle$.

When SWUS-A is used in the S-DWBF algorithm, the resulting algorithm is called the S-DWBF-A
algorithm which at the $l$th iteration updates the CN index set $\mathcal{G}$ with $\mathcal{G}_{\text{A}}^{(l)}$.

Since only a few VNs received updated messages from the selected CNs, some $E_n$'s are likely
to remain constant for many iterations (e.g., $v_3$) or even during the whole decoding process.
To spread the updated messages to more VNs, we expand the updated CN set to include both
$\mathcal{G}_{\text{A}}^{(l)}$ and
\begin{equation}\label{eqn:SWUS-B}
    \mathcal{G}_{\text{B}}^{(l)}\triangleq \{m|m\in\mathcal{M}(n),
~n\in\mathcal{U}_{\text{A}}^{(l)}\}
\end{equation}
where $\mathcal{U}_{\text{A}}^{(l)}\triangleq\{n|n\in
\mathcal{N}(m), ~m\in\mathcal{G}_{\text{A}}^{(l-1)}\}$ and call this updating schedule as the
{\it selective weight-updating schedule B} (SWUS-B).
In other words,
the updated messages received by the VNs connected to the CNs in $\mathcal{G}_{\text{A}}^{(l)}$ will also be forwarded to their connecting CNs in the following iteration,
i.e., $\mathcal{G} \leftarrow \mathcal{G}_{\text{A}}^{(l)}\cup\mathcal{G}_{\text{B}}^{(l)}$.
Fig. \ref{fig:update_B} illustrates the expanded updating range by indicating the extra visited CNs
(those whose indices blong to $\mathcal{G}_{\text{B}}^{(l)}$) with the symbol $\boxbox$. Similarly,
when this weight-updating schedule is used, we call the resulting algorithm as the S-DWBF-B algorithm.

\begin{table*}
    \begin{center} \caption{Complexity of weight-updating schedules}\label{TB:CN_comp}
        \begin{tabular}{|c|c|c|c|} \hline
        ~                 & SWUS-A & SWUS-B & FWUS\\ \hline
        Number of Visited CNs& $d_v$            & $\min\{d_v(d_v-1)(d_c-1)+2d_v, M \}$ & $M$\\ \hline
        Number of Visited VNs& $\min\{d_v(d_c-1)+1, N\}$   & $\min\{d_v(d_v-1)(d_c-1)^2+2[d_v(d_c-1)+1], N\}$ & $N$ \\ \hline
        \end{tabular}
    \end{center}
\end{table*}

Table \ref{TB:CN_comp} lists the average number of visited CNs (to update checksum weights) and
VNs (to compute FF) per iteration per flipped bit or FF inversion for the different weight
updating schedules. As expected, the SWUS-A/B need much less computational complexity and this
reduction is more impressive when $d_c$ and $d_v$ are small. Note that for the expanded schedule
SWUS-B, the VNs in $\mathcal{U}_{\text{A}}^{(l)}$ may be linked to a common set of CNs and
$\mathcal{G}_{\text{A}}^{(l)}\cap\mathcal{G}_{\text{B}}^{(l)}$ can be nonempty if the code graph
has short cycles. As a result, the practical average numbers of the
visited CNs and VNs for SWUS-B is much less than those shown in Table \ref{TB:CN_comp} which
assumes a cycle-free code. However, it can be shown that when the code girth is larger than 8
(10), the actual average visited CN (VN) number is equal to that of a cycle-free code.
\subsection{Performance of S-DWBF Decoding Algorithms}\label{subsection:sDWBF_performance}
\begin{figure}
    \begin{center}
        \epsfxsize=3.6in
        \epsffile{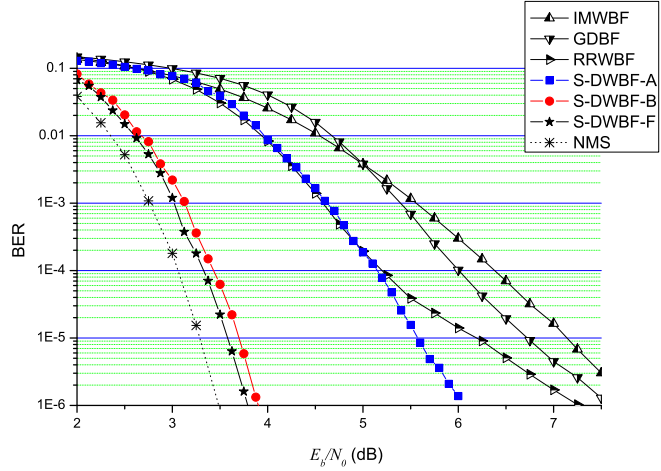}
        \caption{\label{fig:A_M8_SGL_SNRvsBER}BER performance of several single-bit BF (S-BF) decoders
        as a function of SNR ($E_b/N_0$) for Code 1.}
    \end{center}
\end{figure}

\begin{figure}
    \begin{center}
        \epsfxsize=3.6in
        \epsffile{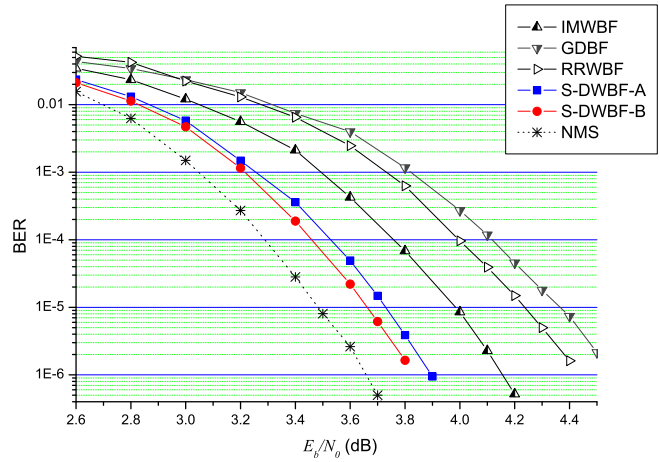}
        \caption{\label{fig:A_EG_SGL_SNRvsBER}BER performance of several S-BF decoder
        as a function of SNR for Code 2.}
    \end{center}
\end{figure}

We apply the known single-bit BF algorithms and S-DWBF-A/B/F algorithms to decode two regular LDPC
codes and present their performance in Figs. \ref{fig:A_M8_SGL_SNRvsBER} and \ref{fig:A_EG_SGL_SNRvsBER}.
The first code, MacKay's $(816,272)(4,6)$ rate-0.333 LDPC code (816.1A4.845 \cite{MacKayWeb}), is a typical
low-rate, low-degree Gallager code with no special structure. In contrast, the second code, the $(1023,781)
(32,32)$ rate-0.763 EG-LDPC code, is a high-rate, high-degree code whose performance has been evaluated in
several WBF-related works \cite{IMWBF}, \cite{AMBF}, \cite{PWBF}, and \cite{IPWBF}. For convenience, we refer
to the above low- and high-rate codes as Code 1 and 2, respectively. It has been shown by simulations \cite{RyanLin}
that BF decoding is very effective in decoding (high-rate, high-degree) EG-LDPC codes. It would be interesting
to examine the BF algorithms in decoding low-rate, low-degree codes. The performance of the normalized min-sum
(NMS) algorithm \cite{NMSA} is also given for reference purpose.

Fig. \ref{fig:A_M8_SGL_SNRvsBER} shows the BER performance of Code 1 with $l_{\text{max}}=150$.
For the S-DWBF-A, S-DWBF-B, and S-DWBF-F decoders, the numerically-optimized $\alpha_2$ values are
0.68, 0.44, and 0.35 whereas for the IMWBF decoder, we found $\alpha_1=0.2$. Our extensive
simulation concluded that the optimal reliability threshold $\eta$ in (\ref{eqn:clipper}) is close
to $0$, whence $\eta=0$ is used for all DWBF algorithms. At BER=$10^{-5}$, we observe that the
S-DWBF-B and S-DWBF-F algorithms have 2.5 dB and 2.6 dB gains against the RRWBF algorithm; the
simple S-DWBF-A algorithm achieves a much smaller 0.7 dB gain as it limit its weight update
to a very limited range.
The performance of Code 2 with $l_{max}=50$ is shown in Fig. \ref{fig:A_EG_SGL_SNRvsBER}. Unlike
Code 1, Code 2 has a much higher $d_v$, and the check reliability part of the GDBF algorithm's FF,
$-\sum_{m\in\mathcal{M}(n)}(1-2s_m)$, thus dominates the FF value after a few iterations
and its performance is similar to that of Gallager's BF algorithm, especially when SNR is high.
To improve its performance we insert a damping factor $\alpha_3$ so that (\ref{eqn:FF-GDBF})
becomes
\begin{eqnarray}\label{eqn:FF-GDBF_alpha}
    E_n=- y_n(1-2\hat{u}_n)-\alpha_3\sum_{m\in\mathcal{M}(n)}(1-2s_m).
\end{eqnarray}
This modification multiplies the second summation of (\ref{eqn:FF-GDBF}) by $\alpha_3$, which is analogous to the
Lagrange multiplier in (the checksum) constrained optimization and when $\alpha_3=1$, (\ref{eqn:FF-GDBF_alpha})
degenerates to (\ref{eqn:FF-GDBF}).
The optimal $\alpha_3$ for (\ref{eqn:FF-GDBF_alpha}) is close to $1/17$ for Code 2. Referring to (\ref{eqn:FF-general}),
the IMWBF algorithm uses $\alpha_1=1.8$ and the S-DWBF-A (B) decoder uses $\alpha_2=0.33$ (0.12) in (\ref{eqn:FF-DWBF}).
Due to the high VN/CN degrees of Code 2, almost all CNs are updated by S-DWBF-B algorithm after
two or three iterations, yielding performance similar to that of the S-DWBF-F decoder. The same figure
shows that the S-DWBF-A decoder provides about 0.25 dB performance gain with respect to the IMWBF
decoder at BER $=10^{-5}$ and the S-DWBF-B algorithm offers additional 0.1 dB gain.
\begin{table}
        \begin{center} \caption{Average number of visited CNs in S-DWBF algorithms}\label{TB:S-DWBF_CNupdates}
            \begin{tabular}{|c|c|c|} \hline
            \multicolumn{3}{|c|}{Code 1 ($M=544$); SNR = 4 dB}\\ \hline
            Iteration & S-DWBF-A & S-DWBF-B \\ \hline
            10        & 15.7     & 137.2    \\ \hline
            30        & 10.1     &  88.7     \\ \hline
            50        & 8.3      &  76.4     \\ \hline
            \multicolumn{3}{|c|}{Code 2 ($M=1023$); SNR = 3.4 dB}\\ \hline
            Iteration & S-DWBF-A & S-DWBF-B \\ \hline
            5         & 105.3    & 1023   \\ \hline
            10        & 56.3     & 1023   \\ \hline
            20        & 38.6     & 1023   \\ \hline
            \end{tabular}

        \end{center}
\end{table}

Table \ref{TB:S-DWBF_CNupdates} presents the average number of CNs visited (in the associated ETFGs) by
different schedules for Code 1 and Code 2
at selected iterations.
Although the S-DWBF-F algorithm has the best BER performance among the single-bit
algorithms, it requires higher computational complexity. By contrast, the S-DWBF-A/B algorithms provide
trade-offs between complexity and performance. Furthermore, for both selective weight-updating schedules, the
number of visited CNs decreases when one proceeds with more iterations as the numbers of flipped bits
and reliability-inverted VNs decreases.
\begin{figure}
    \begin{center}
        \epsfxsize=3.6in
        \epsffile{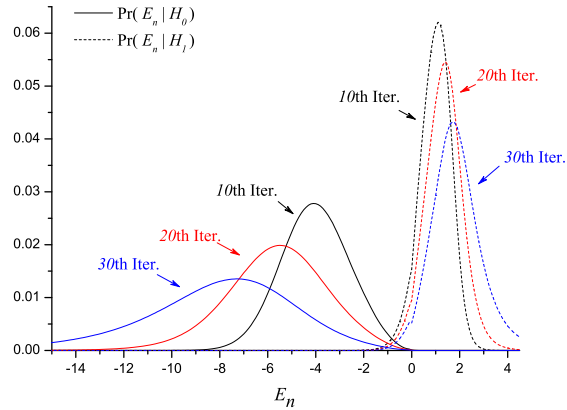}
        \caption{\label{fig:single-DWBF_En}Conditional FF distributions for the S-DWBF-F algorithm in decoding
        Code 1, SNR = 4 dB.}
    \end{center}
\end{figure}
We plot the FF value distributions for the S-DWBF-F algorithm in Fig. \ref{fig:single-DWBF_En}.
In contrast to Fig. 1, where the separation between the two conditional pdfs exhibits little
variation, our DWBF algorithm is able to pull $f(E_n|H_0)$ away from
$f(E_n|H_1)$ as the decoding process evolves. Since the reliability of a decoder's decisions
based on $E_n$ depends on the separation (distance) between the two pdfs, the improved separation
is certainly welcome. As mentioned before, we use $\eta=0$ in (\ref{eqn:clipper}) for all S-DWBF
algorithms. Although the optimal clipping threshold is unknown. Fig. \ref{fig:single-DWBF_En} does
convince us that $0$ is a valid and convenient choice and the FF with the proposed dynamic checksum
weighting does give a much better VN reliability reference.


\section{New Flipped-Bit Selection Methods and Multi-Bit BF Decoding Algorithms}\label{section:multi-DWBF}
Multi-bit BF decoding algorithms were developed to accelerate the convergence performance.
Most of these algorithms use the simple threshold comparison FBS rule (\ref{FBS-Multi}) discussed
in Section \ref{section:Preliminaries} \cite{GDBF}, \cite{ATBF}, \cite{AWBF}, \cite{MGDBF2}--\cite{RNGDBF}.
Even if the threshold $\Delta$ has been optimized by numerical experiments, we find that it is
necessary to add the option $\mathcal{B}=\{n|n=\arg\max_iE_i\}$ in case $\mathcal{B}=\emptyset$
whose occurrence probability during a decoding process is nonzero. In fact, simulation results indicate
that, depending on SNR and the decoding/schedule algorithms used it can be higher than 0.2 for Code 1.
When this option is included in FB decision, the resulting FBS rule is called {\bf Algorithm}
\ref{alg:M1FBS} or the M1-FBS rule.

\begin{algorithm}[H]
    \caption{Flipped Bit Selection Rule 1 (M1-FBS)}\label{alg:M1FBS}
    \textbf{Step} 1 Find $\mathcal{B}=\{n|E_n\geq\Delta\}$. If $\mathcal{B}\neq \emptyset$, stop; otherwise, proceed to \textbf{Step} 2.\\
    \textbf{Step} 2 Find $\mathcal{B}=\{n|n=\arg\max_iE_i\}$
\end{algorithm}

Recall that the PWBF algorithm uses the FF of the IMWBF algorithm as the VN reliability metric with the
FB set (\ref{FBS-PWBF}) determined by the FS count $F_n$ of (\ref{eqn:FScnt}). That is, a CN sends a flip
signal to its most unreliable linked VN only and the VNs which receive sufficient number of reliability
warnings (flip signals) shall be flipped. It turns out that, with this extra filtering of VN-to-CN
messages ($E_n$'s) and selective CN-to-VN flip signal passing, the PWBF algorithm is able to
outperform the IMWBF decoder in both convergence rate and error rate \cite{PWBF}. This performance
gain motivates us to ponder if a more elaborated flipping decision strategy that uses more information
can bring about further performance improvement for the BF decoders using either
the proposed FF (\ref{eqn:FF-DWBF}) or other FFs discussed in Section \ref{section:Preliminaries}.

\subsection{Flipping Intensity}\label{subsection:IDW_FI}
Let $U_n\triangleq\sum_{m\in\mathcal{M}(n)}s_m$, $\mu_m\triangleq\max_{n\in\mathcal{N}(m)}U_{n}$,
$\lambda_m\triangleq\arg\max_{n\in\mathcal{N}(m)}E_{n}$,
and $\mathcal{M}'(n)=\{m|m\in \mathcal{M}(n),\lambda_m=n\}$.
With these notations, we define the \emph{flipping intensity} (FI) of (received by) $v_n$ as
\begin{eqnarray}\label{eqn:FI-UCN}
    \tilde{F}_n&=&\sum_{m\in \mathcal{M}'(n)}\theta_0s_m\delta(U_{\lambda_m}-\mu_m)+\theta_1s_m[1-\delta(U_{\lambda_m}-\mu_m)],
\end{eqnarray}
where $\theta_0>\theta_1\geq 0$ and $\delta(x)$ is the Kronecker delta function. For simplicity,
both $\theta_0$ and $\theta_1$ are confined to be integers so that FI is integer-valued. The above
definition implicitly implies that $\tilde{F}_n=0$ if $\mathcal{M}'(n)=\emptyset$ and only
UCNs have a say in deciding FI. It also implies that a VN has a nonzero FI only if it has
the largest FF value among $\mathcal{N}(m)$ and if it is connected to the largest number
of UCNs among its peers in $\mathcal{N}(m)$, the associated FI should be even higher
($\theta_0 > \theta_1$). In both cases, a UCN $c_m$ will send a non-negative message to
the VN with the highest FF value in the set $\mathcal{N}(m)$. However, if $c_m$ is a passed
CN (PCN) ($s_m=0$) and $d_c$ is small, it often implies that the tentative decisions of
its linked VNs are all correct. Hence if the flipped bits are to be selected by checking
whether the associated FI is greater than a threshold, $v_{\lambda_m}$ should have a
smaller probability of being chosen. This can be done by having the PCN send a drag message
$\theta_2(s_m-1)$. But if there is doubt that $c_m$ is connected to even incorrect bit
decisions, the PCN has better not sending such a message. We decide that this is likely
to be the case if $U_{\lambda_m}\neq\mu_m$ for this inequality means that at least
one VN in $\mathcal{N}(m)$ has more connected UCNs than $v_{\lambda_m}$. With UCNs and
PCNs contributing opposite signals, we modify (\ref{eqn:FI-UCN}) for all $n,~0 \leq n < N$
as
\begin{eqnarray}\label{eqn:FI}
\tilde{F}_n&=&\sum_{m\in \mathcal{M}'(n)}[\theta_2(s_m-1)+\theta_0s_m]\delta(U_{\lambda_m}-\mu_m)+\theta_1s_m[1-\delta(U_{\lambda_m}-\mu_m)],
\end{eqnarray}
where $\theta_2 \leq \theta_0$ is a nonnegative integer. On the other hand, when $d_c$ is large, it is
less likely that $s_m=0$ automatically implies correct decisions on all its linked bits and
we thus stick to (\ref{eqn:FI-UCN}), having no PCN to contribute to FI.
Although the thresholds $\theta_i$'s can be any nonnegative real numbers,
to simplify implementation, we let $\theta_i$'s be nonnegative integers
such that the FI is integer-valued.

\subsection{Flipped-Bit Selection Rule}\label{subsection:New_B_building}
A simple FI-based FBS rule is to flip the bits in the FB set $\mathcal{B}=\{n|\tilde{F}_n
\geq\Delta_{\text{FI}}\}$. But the optimal threshold $\Delta_{\text{FI}}$ is not easy to determine
especially for a code with low VN degree. A smaller threshold may cause incorrect flipping decisions
while a large threshold tends to slow down the convergence or even cause decoding failure as no VN
meets the the flipping requirement. To overcome this dilemma, we select a relative high FI threshold
and use the FB set $\mathcal{B}=\{n|\tilde{F}_n \geq \Delta_{\text{FI}}\}$ if it is nonempty. Otherwise,
$\mathcal{B}=\{n|U_n=\max_{i\in\mathcal{T}}U_i\}$ where $\mathcal{T}\triangleq \{n|\tilde{F}_n=\max_{j}
\tilde{F}_j\}$.
We summarize below the new FBS rule as \textbf{Algorithm} \ref{alg:M2FBS} which, for
convenience of reference, is called the M2-FBS rule.
\begin{algorithm}[H]
    \caption{Flipped Bit Selection Rule 2 (M2-FBS)}\label{alg:M2FBS}
    \textbf{Step} 1 For $n=0,1,\ldots,N-1$, compute $\tilde{F}_{n}$ by (\ref{eqn:FI-UCN}) or (\ref{eqn:FI}).\\
    \textbf{Step} 2 Find $\mathcal{B}=\{n|\tilde{F}_n\geq\Delta_{\text{FI}}\}$. If $\mathcal{B}\neq \emptyset$, stop; otherwise, proceed to \textbf{Step} 3.\\
    \textbf{Step} 3 Update $\mathcal{T}$ and find $\displaystyle\mathcal{B}=\{n|U_n=\max_{i\in\mathcal{T}}U_{i}\}$.
\end{algorithm}

Note this FBS rule is independent of the FF and can be used
in conjunction with different FFs no matter whether the checksum weights are constant or not.

Loop-detection/breaking procedures can be included in our FBS algorithm if necessary.
The loop detection scheme used \cite{AMBF} is an appropriate choice.
When a loop is detected, we generate a disturbance on the tentative decoded sequence by switching to the FB set
\begin{eqnarray}\label{eqn:FBS-LoopBreak}
    \mathcal{B}=\{n| U_n=\max_{i}U_{i}\}.
\end{eqnarray}

\subsection{Numerical Results}\label{subsection:FBS_performance}
Different combinations of the FBS rule, the FF, and the weight-updating schedule,
used lead to different decoding algorithms.
The error-rate performance and decoding speed of various decoders are presented in this subsection.
\subsubsection{Abbreviations}
For convenience of reference, we adopt a systematic labeling scheme similar to that used in
Section \ref{subsection:sDWBF_performance} to describe a decoding method. We denote a decoder
by groups of capital letters separated by hyphens, specifying respectively the FBS rule, the FF
and the weight-updating schedule used with the 3-field form, {\it FBS rule-FF-weight updating schedule}.
That is, the first filed is used to indicate if single (S) or multiple (M) bits
are to be flipped in an iteration and, for the latter case, if the simple FF based (M1) or the
more complicated FI-based (M2) FBS rule is adopted. The second field contains the abbreviation
of the known or proposed algorithm such as IMWBF, GDBF or DWBF whose FF is used. The third field
tells whether a selective (A or B) or the full (F) weight-updating schedule is used.
Since only the DWBF algorithms need to update checksum weights, the third field is omitted for
non-DWBF based decoders. Hence, M1-DWBF-A represents the decoder that uses the M1-FBS rule,
the DWBF FF, and SWUS-A, and M2-IMWBF(-GDBF) denotes the decoder that uses the M2-FBS rule and
the IMWBF (GDBF) algorithm's FF. For known constant weight algorithms without FBS modification and
SWUS, we keep conventional abbreviations like AMWBF, IPWBF, and HGDBF only.
\subsubsection{Parameter values used}
\begin{table}
    \begin{center} \caption{Simulation parameter settings}\label{TB:parameters}
        \begin{tabular}{|c|c|c|}
          \hline
          Algorithm & Code 1 & Code 2 \\ \hline \hline
          HGDBF & $\alpha_3=1$ & $\alpha_3=1/17$ \\ \hline
          AMWBF & $\alpha_1=0.2$ & $\alpha_1=1.8$ \\ \hline
          IPWBF & $\alpha_1=0.2,\Delta_{\text{FS}}=1$ & $\alpha_1=1.8,\Delta_{\text{FS}}=10$ \\ \hline
          M2-IMWBF & $\alpha_1=0.2,\Delta_{\text{FI}}=5$ & $\alpha_1=3.2,\Delta_{\text{FI}}=16$ \\ \hline
          M2-GDBF & $\alpha_3=1,\Delta_{\text{FI}}=1$ & $\alpha_3=1/17,\Delta_{\text{FI}}=10$ \\ \hline
          M1-DWBF-A & $\alpha_2=0.7,\Delta=0$ & $\alpha_2=0.33,\Delta=0$ \\ \hline
          M1-DWBF-B & $\alpha_2=0.35,\Delta=0$ & $\alpha_2=0.3,\Delta=0$ \\ \hline
          M2-DWBF-A & $\alpha_2=0.58,\Delta_{\text{FI}}=1$ & $\alpha_2=0.33,\Delta_{\text{FI}}=4$ \\ \hline
          M2-DWBF-B & $\alpha_2=0.35,\Delta_{\text{FI}}=1$ & $\alpha_2=0.3,\Delta_{\text{FI}}=1$ \\
          \hline
        \end{tabular}
    \end{center}
\end{table}

For the decoders based on M2-FBS rule, we use the FI weights $\theta_0=3,~\theta_1=2$, and
$\theta_2=1$. Other major parameter values for different multi-bit BF algorithms are listed
in Table \ref{TB:parameters}. The remaining parameters needed for the IPWBF algorithm follow
those suggested in \cite{IPWBF}, and those associated with the escaping process in the HGDBF
algorithm are also optimized. For simplicity, the FF clipping threshold $\eta$ in (\ref{eqn:clipper})
and the threshold $\Delta$ used in (\ref{FBS-Multi}) in M1-DWBF-A/B algorithms are set to zero.

Note that the parameters associated with a decoder are correlated, i.e., if the optimal value of a parameter
is dependent on other parameters' values used, although the correlation may not be very high. Hence, we try
to jointly optimize these parameters to minimize the converged error rate. Furthermore, simulation results
indicate that the optimal parameter values are insensitive to SNR.

\subsubsection{BER and FER performance}
\begin{figure}
	\begin{center}
		\epsfxsize=3.6in \epsffile{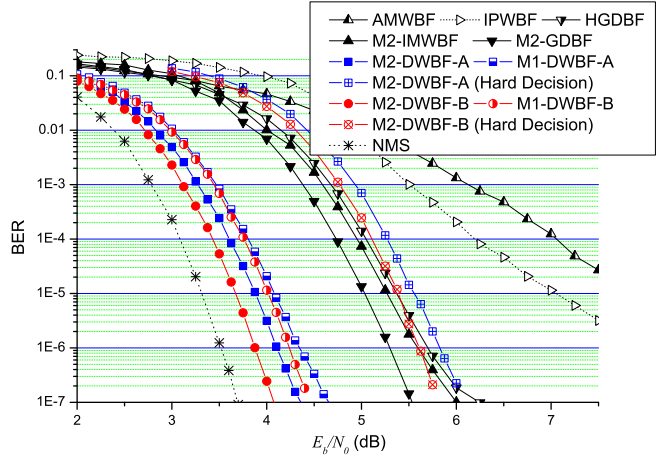}\caption{
		\label{fig:B_M8_MLT_SNRvsBER}BER performance of various multi-bit BF (M-BF) decoding algorithms
        as a function of SNR for Code 1.}
	\end{center}
\end{figure}
Fig. \ref{fig:B_M8_MLT_SNRvsBER} shows the BER performance of different multi-bit BF algorithms for
Code 1 when $l_{\text{max}}=50$. The effectiveness of the M2-FBS rule can also be verified by
comparing the required $E_b/N_0$ for BER $=10^{-5}$: the M2-IMWBF decoder outperforms the IPWBF
decoder by approximately 1.7 dB and the M2-GDBF algorithm has a 0.4 dB gain over the HGDBF
algorithm.
The DWBF algorithms yields better BER performance even with the simple M1 rule and,
when the M2 rule is used, its performance becomes closer (0.4 dB) to that provided by the NMS algorithm.
\begin{figure}
	\begin{center}
		\epsfxsize=3.6in \epsffile{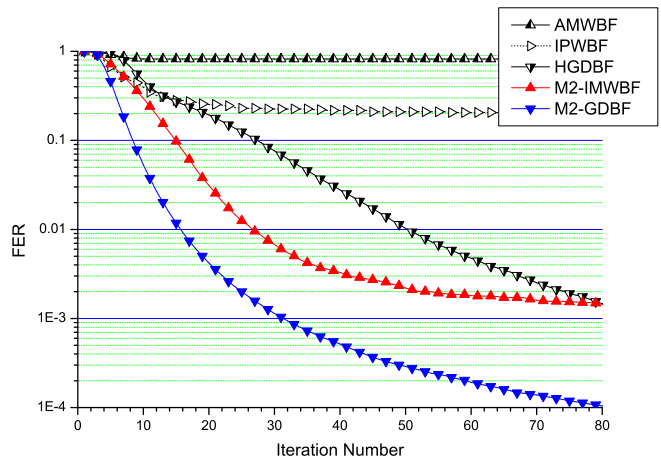}\caption{
		\label{fig:C_M8_MLT_ITEvsFER_PBS}Frame error rate (FER) convergence performance of various M-BF decoding algorithms
        using conventional or M2-FBS rule; Code 1, SNR = 5 dB.}
	\end{center}
\end{figure}
\begin{figure}
	\begin{center}
		\epsfxsize=3.6in \epsffile{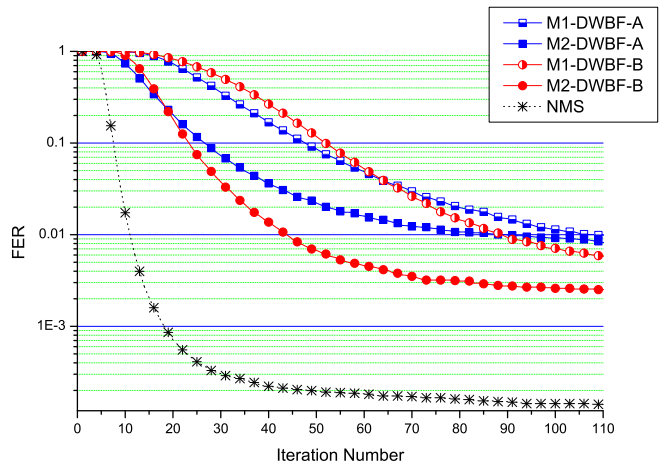}\caption{
		\label{fig:C_M8_MLT_ITEvsFER_DMS}FER convergence performance of multi-bit DWBF (M-DWBF) decoding algorithms; Code 1, SNR = 3.25 dB.}
	\end{center}
\end{figure}
The convergence behaviors of these algorithms are shown in Figs. \ref{fig:C_M8_MLT_ITEvsFER_PBS}
and \ref{fig:C_M8_MLT_ITEvsFER_DMS}. The results show that the M2 rule gives better BER performance
and, for both the DWBF and M2-GDBF algorithms, the convergence rate is improved as well.

Note that in Figs. \ref{fig:B_M8_MLT_SNRvsBER}-\ref{fig:C_M8_MLT_ITEvsFER_DMS}, loop-detecting/breaking
schemes are activated for all but the M2-DWBF-B algorithm. In general, loops are much less likely to
occur in a DWBF decoder than in a static CN weight decoder. When the FWUS or SWUS-B is used to decode
Code 2, our simulations detect no loop for both codes whence there is no need for a loop breaker. This
is because the time-varying checksum weights of the DWBF algorithm and wider message magnitude propagation
ranges of the FWUS or SWUS-B schedule have made the BF decision related variables, $E_n, U_n, \mu_m$, and
$\tilde{F}_n$, to have much larger dynamic ranges; see also Figs. \ref{fig:En-IMF} and \ref{fig:single-DWBF_En}.
\begin{figure}
	\begin{center}
		\epsfxsize=3.6in \epsffile{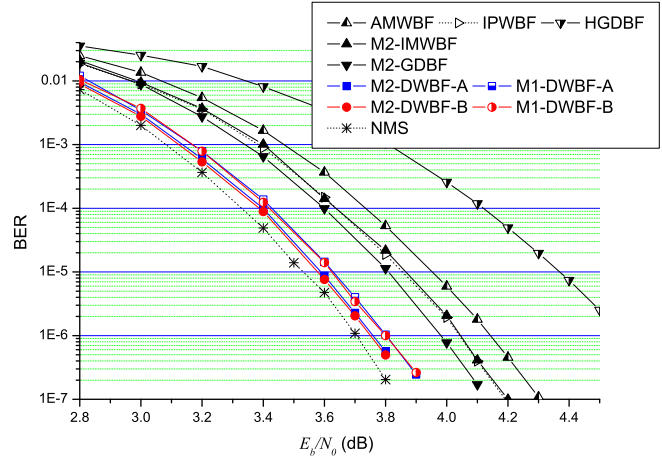}\caption{
		\label{fig:B_EG_MLT_SNRvsBER}BER performance of various M-BF decoding algorithms as a function of SNR for Code 2.}
	\end{center}
\end{figure}
\begin{figure}
	\begin{center}
		\epsfxsize=3.6in \epsffile{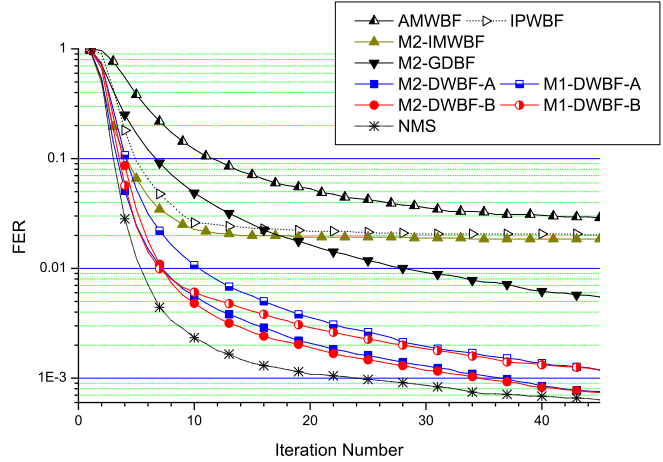}\caption{
		\label{fig:C_EG_MLT_ITEvsFER}FER convergence performance of several M-BF decoding algorithms; Code 2, SNR = 3.4 dB.}
	\end{center}
\end{figure}

The BER (with $l_{\text{max}}=20$) and frame error rate (FER) convergence performance of various
multi-bit BF decoders for Code 2 are respectively presented in Figs. \ref{fig:B_EG_MLT_SNRvsBER}
and \ref{fig:C_EG_MLT_ITEvsFER}. By comparing the two sets of BER curves, M2-GDBF versus HGDBF
and M2-IMWBF versus AMWBF, we verify the effectiveness of the new FBS (M2) rule. Although the
M2-IMWBF algorithm yields the same converged BER as that of the IPWBF decoder for this code,
it gives better FER performance in the first few iterations. We further notice that the M2-DWBF-A
(B) decoder is superior to the M1-DWBF-A (B) decoder in both BER performance and decoding speed.
The former yields performance very close to that of the NMS algorithm while the latter suffers only about
0.1 dB performance degradation against the NMS decoder at BER=$10^{-5}$. Figs.
\ref{fig:B_EG_MLT_SNRvsBER} and \ref{fig:C_EG_MLT_ITEvsFER} also show that the performance gap
between the M1/M2-DWBF-A and M1/M2-DWBF-B decoders is smaller than that for Code 1. This is due
to the high VN/CN degrees of Code 2: the high VN degree increases the probability that a CN is
visited by the SWUS-A while the high CN degree helps spreading the updated weights to more VNs.

We want to remark that a) only the AMWBF and HGDBF decoders need a loop-breaker in decoding
Code 2 since for the other decoders, loops are rarely detected and b) although the M1-FBS rule
is simpler, our simulations indicate that the M2-FBS rule can significantly reduce the probability
of decoding loops when it is used in conjunction with the DWBF, GDBF, or IMWBF algorithms.
This is particular useful when using conventional FFs to decode low-degree codes.

\subsection{Complexity Analysis}\label{subsection:FBS_complexity}
\begin{table*}
        \begin{center} \caption{Average UCN and visited CN numbers}\label{TB:CNupdates}
       \scalebox{0.8}{
            \begin{tabular}{|c|c|c|c|c|c|c|c|c|c|c|c|c|c|c|} \hline
            \multicolumn{15}{|c|}{Code 1 ($M=544$); SNR = 3.25 dB}\\ \hline
            \multirow{2}{*}{Iteration ($l$)}
                &IPWBF&\multicolumn{2}{c|}{M2-IMWBF}&M2-GDBF&\multicolumn{2}{c|}{M1-DWBF-A}&\multicolumn{2}{c|}{M1-DWBF-B}&\multicolumn{3}{c|}{M2-DWBF-A}&\multicolumn{3}{c|}{M2-DWBF-B}\\ \cline{2-15}
                &$M_{\text{1}}^{(l)}$&$M$&$P_{\text{S3}}^{(l)}$&$M$&$M_{\text{WU}}^{(l)}$&$P_{\text{S2}}^{(l)}$&$M_{\text{WU}}^{(l)}$&$P_{\text{S2}}^{(l)}$
                &$M$&$M_{\text{WU}}^{(l)}$&$N_{\text{FB}}^{(l)}$&$M$&$M_{\text{WU}}^{(l)}$&$N_{\text{FB}}^{(l)}$\\ \hline

            5&91.7&\multirow{4}{*}{544}&0.36&\multirow{4}{*}{544}&115.2&$3.6\times 10^{-5}$&386.7&$3.5\times 10^{-6}$&\multirow{4}{*}{544}&135.2&34.1&\multirow{4}{*}{544}&506.6&17.2\\
            \cline{1-2}\cline{4-4}\cline{6-9}\cline{11-12}\cline{14-15}

            10&78.3&&0.52&&110.6&$0.0035$&162.8&0.0014&&80.1&19.1&&414.2&8.8\\ \cline{1-2}\cline{4-4}\cline{6-9}\cline{11-12}\cline{14-15}

            15&71.0&&0.67&&107.7&$0.0092$&132.1&0.0019&&69.0&16.5&&338.1&6.3\\ \cline{1-2}\cline{4-4}\cline{6-9}\cline{11-12}\cline{14-15}

            20&68.6&&0.65&&96.5&$0.0099$&131.3&0.0012&&66.9&15.8&&328.7&5.9\\ \hline

            \multicolumn{15}{|c|}{Code 2 ($M=1023$); SNR = 3.4 dB}\\ \hline
            \multirow{2}{*}{Iteration ($l$)}
                &IPWBF&\multicolumn{2}{c|}{M2-IMWBF}&M2-GDBF&\multicolumn{2}{c|}{M1-DWBF-A}&\multicolumn{2}{c|}{M1-DWBF-B}&\multicolumn{3}{c|}{M2-DWBF-A}&\multicolumn{3}{c|}{M2-DWBF-B}\\ \cline{2-15}
                &$M_{\text{1}}^{(l)}$&\multicolumn{2}{c|}{$M_{\text{1}}^{(l)}$}&$M_{\text{1}}^{(l)}$&\multicolumn{2}{c|}{$M_{\text{WU}}^{(l)}$}&\multicolumn{2}{c|}{$M_{\text{WU}}^{(l)}$}
                &$M_{\text{1}}^{(l)}$&$M_{\text{WU,0}}^{(l)}$&$N_{\text{FB}}^{(l)}$&$M_{\text{1}}^{(l)}$&$M_{\text{WU,0}}^{(l)}$&$N_{\text{FB}}^{(l)}$\\ \hline

            3&172.1&\multicolumn{2}{c|}{217.2}&326.2&\multicolumn{2}{c|}{267.3}&\multicolumn{2}{c|}{1023}&281.6&382.4&18.3&332.4&690.6&24.6\\ \hline

            5&198.3&\multicolumn{2}{c|}{287.7}&351.9&\multicolumn{2}{c|}{374.9}&\multicolumn{2}{c|}{1023}&318.5&386.9&26.1&322.7&700.3&27.0\\ \hline

            10&329.4&\multicolumn{2}{c|}{360.2}&407.2&\multicolumn{2}{c|}{588.7}&\multicolumn{2}{c|}{1023}&425.8&415.4&46.4&436.5&586.5&50.7\\ \hline

            15&353.6&\multicolumn{2}{c|}{378.4}&414.6&\multicolumn{2}{c|}{644.5}&\multicolumn{2}{c|}{1023}&438.1&421.0&48.2&445.0&577.0&59.5\\ \hline
            \end{tabular}
            }
        \end{center}
\end{table*}

\begin{table*}
        \begin{center} \caption{Computational complexity for various decoding algorithms (C1: Code 1, C2: Code 2)}\label{TB:TTL_Comp}
        \scalebox{0.72}{
            \begin{tabular}{|c|c|c|c|c|c|c|} \hline
            Operation          &HGDBF &IPWBF                              & M2-GDBF/IMWBF      & M1-DWBF-A/B
            & M2-DWBF-A/B        & NMS\\ \hline

            \multirow{2}{*}{Integer Additions}     & \multirow{2}{*}{$0$}  & \multirow{2}{*}{$M_{\text{1}}^{(l)}$} & C1: $M~~~$ & \multirow{2}{*}{0}
            & C1: $M~~~$ & \multirow{2}{*}{0} \\

            &&&C2: $M_{\text{1}}^{(l)}$&&C2: $M_{\text{1}}^{(l)}$& \\ \cline{1-7}

            Real Number Additions& $0$  & $0$                               & $0$                & 0
            & 0                  & $N d_v$ \\ \hline

            \multirow{4}{*}{Integer Comparisons}    &  \multirow{4}{*}{$0$}  &  \multirow{4}{*}{$N$}&C1 (M2-IMWBF): $M (d_c-1)+N$& \multirow{4}{*}{0}
            && \multirow{4}{*}{0}\\

            &&&$~~~~~~~~~~~~+P_{\text{S3}}^{(l)}N$&&C1: $M (d_c-1)+N$~~~~&\\

            &&&~~~~~~~C1 (M2-GDBF): $M (d_c-1)+N~~~~~~~$&&C2: $M_{\text{1}}^{(l)} (d_c-1)+N$&\\

            &&&~~~~~~~~~~~~~~~~~~C2: $M_{\text{1}}^{(l)} (d_c-1)+N$&&& \\

            \cline{1-7}

            \multirow{4}{*}{Real Number Comparisons} & \multirow{4}{*}{$N$}  & \multirow{4}{*}{$M_{\text{1}}^{(l)}(d_c-1)$}&&
            C1: $M_{\text{WU}}^{(l)}(2d_c-3)$
            &C1: $M (d_c-1)+M_{\text{WU}}^{(l)}(d_c-2)$  & \multirow{4}{*}{$M(2d_c-3)$} \\

            &&&C1: $M(d_c-1)~~~$&~~~~~$+P_{\text{S2}}^{(l)}(N-1)$&$+N_{\text{FB}}^{(l)} d_v$~~~~~~~~~~~~~~~~~~& \\

            &&&C2: $M_{\text{1}}^{(l)}(d_c-1)$& C2: $M_{\text{WU}}^{(l)}(2d_c-3)$&C2: $(M_{\text{1}}^{(l)}+M_{\text{WU,0}}^{(l)})(2d_c-3)$~~~& \\

            &&&&&$+N_{\text{FB}}^{(l)} d_v$~~~~~~~~~~~~~~~~~~& \\ \hline

            \end{tabular}
            }
        \end{center}
\end{table*}

\begin{table*}
    \begin{center} \caption{Averaged overall complexity ($\times 10^{3}$) per frame for achieving FER=$10^{-3}$ (Real: real comparison or addition; Int: integer comparison or addition)}\label{TB:DWBF_Comp}
    \scalebox{1.1}{
        \begin{tabular}{|c|c|c|c|c|c|c|c|c|c|c|c|c|}
          \hline
          \multicolumn{13}{|c|}{Code 1}\\ \hline
          \multirow{2}{*}{SNR} & \multicolumn{2}{c|}{M1-DWBF-A} & \multicolumn{2}{c|}{M1-DWBF-B} & \multicolumn{3}{c|}{M2-DWBF-A} & \multicolumn{3}{c|}{M2-DWBF-B}  & \multicolumn{2}{c|}{NMSA} \\ \cline{2-13}

          &$l_{\text{max}}$&Real&$l_{\text{max}}$&Real&$l_{\text{max}}$&Real&Int.&$l_{\text{max}}$&Real&Int.&$l_{\text{max}}$&Real\\ \hline

          3.5 dB  &132&19.1&101&46.4&109&41.7&51.1&47&64.1&54.4&12&44.2 \\ \hline

          3.625 dB&89 &15.3&76 &41.7&66 &36.2&49.6&35&57.9&49.5&10&41.6 \\ \hline

          3.75 dB &71 &12.5&64 &38.1&42 &32.1&45.5&28&52.8&45.4&9&39.2 \\ \hline
          \multicolumn{13}{|c|}{Code 2}\\ \hline
          \multirow{2}{*}{SNR} & \multicolumn{2}{c|}{M1-DWBF-A} & \multicolumn{2}{c|}{M1-DWBF-B} & \multicolumn{3}{c|}{M2-DWBF-A} & \multicolumn{3}{c|}{M2-DWBF-B}  & \multicolumn{2}{c|}{NMSA} \\ \cline{2-13}

          &$l_{\text{max}}$&Real&$l_{\text{max}}$&Real&$l_{\text{max}}$&Real&Int.&$l_{\text{max}}$&Real&Int.&$l_{\text{max}}$&Real\\ \hline

          3.4 dB&49&83.5&46&172.5&35&176.4&66.7&34&269.1&80.3&22&255.9 \\ \hline

          3.6 dB&11&59.6&8 &135.2&8 &130.7&50.4&7 &210.7&62.3&6 &218.2 \\ \hline

          3.7 dB&7 &53.3&6 &123.0&6 &115.4&44.9&6 &189.3&55.7&5 &206.8 \\ \hline
        \end{tabular}
        }
    \end{center}
\end{table*}

Besides the syndrome computing, which is the same for all algorithms, the computational
complexity of a BF decoding algorithm consists mainly of three parts: i) FF update, ii)
flipped bits selection, and iii) weight/message update. Once new CN messages, $-w_{mn}
(1-2s_m)$ in (\ref{eqn:FF-general}) or $-r_{mn}(1-2s_m)$ in (\ref{eqn:FF-DWBF}), are
available, the FF update is just adding all returned CN messages and $-\alpha_1|y_n|$
or $-y_n(1-2\hat{u}_n)$. There is little difference among the BF decoders in FF computing.
The only exception is that used by GDBF algorithms, both (\ref{eqn:FF-GDBF}) and
(\ref{eqn:FF-GDBF_alpha}) require integer additions only. iii) is needed for DWBF algorithms
but not other BF algorithms which require only a sign change. Therefore, in the next three
subsections, we consider ii) first, followed by the discussion of iii), the extra complexity
requirement for DWBF algorithms, and finally compare the combined computational complexity of ii)
and iii). The additional complexity such as that associated with a loop-breaking scheme,
is addressed at the end of this section as well. Since most algorithms, except those
using the M2-FBS rule, which need additional integer operations and memory for storing
the FIs and UCN numbers, require approximately the same storage space, we discuss only the
computational complexity.
\subsubsection{FBS complexity}
The HGDBF decoder needs only $N$ real comparisons in (\ref{FBS-Multi}) in selecting
the flipped bits. For the IPWBF decoder, $d_c-1$ real comparisons are required to
find the most unreliable connected VN per UCN and a total of $M_{\text{1}}^{(l)}$
integer additions and $N$ integer comparisons are needed to compute the FS and decide
the FB set (\ref{FBS-PWBF}), where $M_{\text{1}}^{(l)}$ is the average UCN number in
the $l$th iteration.

For the M1-FBS rule, the complexity of \textbf{Step} 1 is ignored since we use $\Delta=0$
and a threshold comparison needs a sign-bit check only. The average complexity of \textbf{Step}
2 is $P_{\text{S2}}^{(l)}(N-1)$ real comparisons, where $P_{\text{S2}}^{(l)}$ is the probability that
\textbf{Step} 2 is activated in the $l$th iteration. For the M2-FBS rule (\textbf{Algorithm}
\ref{alg:M2FBS}), \textbf{Step} 1 needs $d_c-1$ real and $d_c-1$ integer comparisons
per CN in finding $\lambda_m$ and checking if $v_{\lambda_m}$ has the most connected UCNs.
Each CN has to send an integer-valued message, $\theta_0$, $\theta_1$, or $-\theta_2$,
to one of its connected VNs, implying an integer addition in (\ref{eqn:FI}) or
(\ref{eqn:FI-UCN}). Since the former involves both UCNs and PCNs while the latter
involves only UCNs, all $M$ CNs have to perform all the above operations when decoding
Code 1 in contrast to $M_{\text{1}}^{(l)}$  CNs for Code 2. Moreover, $N$ integer
comparisons are required in \textbf{Step} 2 for deciding the FB set. The average complexity
of \textbf{Step} 3 is approximately equal to $P_{\text{S3}}^{(l)}N$ integer comparisons, where
$P_{\text{S3}}^{(l)}$ is the probability that \textbf{Step} 3 is activated at the $l$th iteration.

\subsubsection{Weight update complexity}
Updating the weights associated with CN $c_m$ in the M1-DWBF-A/B decoders require $2d_c-3$ real
comparisons for finding the indices associated with the smallest and second smallest $-E_n$'s,
$n \in \mathcal{N}(m)$. For M2-DWBF-A/B decoders, however, most of $\lambda_m$'s have been found in
the FBS step, hence only the second smallest ones remain to be found for computing new weights.
A more detailed analysis is given in the next two paragraphs.

We first consider a low CN degree code such as Code 1. After CNs compute $\lambda_m$'s and VNs compute
their FIs via (\ref{eqn:FI}), the flipped bits are decided and flipped. The M2-DWBF-A/B algorithms then
invert the associated FF values (\textbf{Step} 3 of \textbf{Algorithm} \ref{alg:DWBF}) and update the
CN weights (\textbf{Step} 4 of \textbf{Algorithm} \ref{alg:DWBF}). As only a small portion of the visited
CNs are connected to the flipped bits, most visited CNs require only $d_c-2$ real comparisons for finding
the second smallest $-E_n$. For a visited CN that links to flipped bits, we only need to compare the
connected flipped bits' $E_n$'s with the original smallest $-E_n$ to find a new minimum $-E_n$ since
only the flipped bits' $E_n$'s are changed between the bit flipping and weight updating. As a result, for
a visited CN linking to $t$ flipped bits, only additional $t$ real comparisons are required for updating
the smallest $-E_n$ before finding the second smallest $-E_n$. If we denote by $M_{\text{WU}}^{(l)}$ and
$N_{\text{FB}}^{(l)}$ the average numbers of visited CNs and flipped bits at the $l$th iteration,
respectively, we need, on the average,
at most $M_{\text{WU}}^{(l)}(d_c-2)+N_{\text{FB}}^{(l)} d_v$ real comparisons for updating weights,
where $N_{\text{FB}}^{(l)} d_v$ accounts for the sum of all additional ($t$) comparisons.

Decoding a high CN degree code requires $2d_c-3$ real comparisons for updating the weight of a visited PCN
and $d_c-2+t$ for an UCN, as the FI formula (\ref{eqn:FI-UCN}) involves only UCNs. Denote the average numbers
of visited PCNs and UCNs at the $l$th iteration by $M_{\text{WU,0}}^{(l)}$ and $M_{\text{WU,1}}^{(l)}$. We
observe from simulations that almost all UCNs are visited (i.e., $M_{\text{WU,1}}^{(l)}=M_{\text{1}}^{(l)}$).
Hence, M2-DWBF-A/B decoders require an average of $M_{\text{WU,0}}^{(l)}(2d_c-3)$ and at most $M_{\text{1}}^{(l)}
(d_c-2)+N_{\text{FB}}^{(l)}d_v$ real comparisons per iteration for computing new CN weights of visited PCNs
and UCNs, respectively. For both cases we ignore the complexity of the threshold comparison in (\ref{eqn:dynWeight})
since $\eta=0$.

\subsubsection{Complexity Summary}

Table \ref{TB:CNupdates} presents the simulated average numbers of $M_{\text{1}}^{(l)}$,
$M_{\text{WU}}^{(l)}$, $M_{\text{WU,0}}^{(l)}$, $N_{\text{FB}}^{(l)}$, $P_{\text{S2}}^{(l)}$,
and $P_{\text{S3}}^{(l)}$ at selected iterations for the IPWBF, M2-IMWBF/GDBF, M1-DWBF-A/B, and M2-DWBF-A/B algorithms.
Since the simulation results indicate that when decoding Code 2 with the M1-DWBF-A/B decoders,
\textbf{Step} 2 of the M1-FBS rule is {\it never} activated, we list the $P_{S2}^{(l)}$ values for Code 1 only.
Similarly, \textbf{Step} 3 of the M2-FBS rule is needed {\it only if} the M2-IMWBF algorithm is used to
decode Code 1, we thus specify the $P_{S3}^{(l)}$ values for this case only.

Considering both the FBS rules and weight updating, we summarize the computational complexity which includes real/integer
additions and comparisons per iteration for various BF and the NMS algorithms in Table \ref{TB:TTL_Comp}.
As computing the total LLRs in the NMS algorithm requires the same efforts as
that of computing FF values in BF decoders,
only the efforts needed for computing the CN-to-VN
and VN-to-CN messages are listed in the table.

Table \ref{TB:DWBF_Comp} presents the simulated average complexity of various DWBF and the NMS algorithms
for decoding a frame with a target FER of $10^{-3}$ at different SNRs.
As an integer (or real) comparison requires about the same computational complexity as that of an integer
(or real) addition (hardware implementation of comparison can even be simpler than addition). Both are thus
counted equally. We show in this table the average integer and real operations and the maximum iteration
number ($l_{\text{max}}$) needed.

Tables IV-VI and Figs. \ref{fig:B_M8_MLT_SNRvsBER}--\ref{fig:C_EG_MLT_ITEvsFER} provide useful information
for studying tradeoffs between performance, complexity, and convergence rate when combining different FBS rules,
FFs and weight-updating schedules. In particular, Table \ref{TB:DWBF_Comp} shows that when decoding Code 1,
both M1-DWBF-A and M1-DWBF-B algorithms need less computational complexity to achieve FER=$10^{-3}$ than that
needed by the NMS algorithm in higher SNR (say, $> 3.625$ dB) region. The M2-DWBF-A algorithm also needs less
real operations in comparison with the NMS algorithm. For decoding Code 2, the M1-DWBF-A, M1-DWBF-B, and M2-DWBF-A
algorithms need less total (real + integer) operations to achieve the FER requirement while the required iteration
numbers are also comparable to that needed for the NMS algorithm in higher SNR region.
Furthermore, based on Tables \ref{TB:CNupdates}, \ref{TB:TTL_Comp}, and Fig. \ref{fig:B_M8_MLT_SNRvsBER},
we conclude that the M2-IMWBF and M2-GDBF algorithms require far less complexity than that of the
IPWBF algorithm in decoding Code 1.

Among the the decoding algorithms compared in Table \ref{TB:TTL_Comp}, the IPWBF algorithm
uses a simpler FBS operation but it has to perform a delay-handling process in every iteration plus
an initial bootstrapping step. These two extra operations need off-line computing effort in searching
for the corresponding optimal parameter values. They also require additional storage and computational
complexity. Although the HGDBF algorithm does not have to sort the FF values, three real thresholds,
one for the multi-bit flipping mode and two for the escape (loop-breaking) process are required in
its FBS rule, resulting extra off-line search and random variable generations. Other off-line efforts
include the searches for $\alpha_1$ (M2-IMWBF), $\alpha_2$ (M1- and M2-DWBF-A/B), and $\Delta_{\text{FI}}$
($\in [-d_v \theta_2, d_v \theta_0]$). As mentioned before, they must be jointly optimized. For the
M2-GDBF algorithm, only the optimal $\Delta_{\text{FI}}$ has to be found.

Our loop-breaking scheme (\ref{eqn:FBS-LoopBreak}) is simpler than those used by
other decoding algorithm and more effective than the methods used by the IPWBF
and AMWBF algorithms which remove the bit(s) having maximum $F_n$ or $E_n$ from
$\mathcal{B}$; when $\mathcal{B}=\emptyset$, the decoding process will be forced
to terminate after the removal. Instead of reducing $|\mathcal{B}|$, the escape
process (\ref{eqn:FBS-LoopBreak}) and that used by the HGDBF algorithm perturb
the tentative decoded sequence to break a loop. The latter, however, has to
generate Gaussian random variables.

\section{Conclusion}\label{section:conclusion}
We divide a typical BF LDPC code decoding algorithm into three major components, namely
1) VN decision reliability (FF) computing and the associated CN reliability (checksum
weight) update formula; 2) the FBS rule; and 3) the checksum weight-updating schedule.
These three components determine the performance and complexity of a BF decoder. We
develop novel FF and FBS rules to improve the BF decoding performance. On the other hand,
the checksum weight update operation is a complexity concern for the DWBF decoders, 
we propose selective weight-updating schedules to reduce the implementation
complexity with little performance loss.

Different combinations of FF, checksum weight-updating method and schedule, FBS rule, result
in different decoder structures. We simulate the error rate and convergence performance of
various decoders, and the resulting numerical behaviors confirm the effectiveness 
of our new design proposals. We show that the combinations of the new multi-bit FBS rules 
with known BF algorithms achieve significant performance gain especially for a high-rate
code. Detailed complexity analysis on various decoder structures is provided for complexity 
and performance tradeoff studies.
We find that, compared with the NMS algorithm,
the combination of the new FBS rules with our DWBF algorithms require less complexity 
in achieving a target FER if SNR is sufficiently high. 
We also find that the convergence rates
are comparable when decoding a high-rate code.

\end{document}